\documentclass{article}

\usepackage{graphicx}
\usepackage[round]{natbib}
\usepackage{psfig}
\usepackage{epsfig}

\title{
Year ahead prediction of US landfalling hurricane numbers: the
optimal combination of long and short baselines for intense
hurricanes
 }

\author{Stephen Jewson\footnote{\emph{Correspondence email}: \texttt{x@stephenjewson.com}}, Jeremy Penzer and Christopher Casey\\}

\begin{document}
\maketitle

\begin{abstract}
In previous work, we have shown how to combine long and short historical baselines
to make predictions of future hurricane numbers. We now ask: how should
such combinations change if we are interested in predicting the future number
of \emph{intense} hurricanes?
\end{abstract}

\section{Introduction}

We are interested in predicting hurricane numbers well in advance
of the start of the hurricane season.
There are a number of methods that one might consider to try and use to
make such predictions, such as:

\begin{itemize}
    \item Time series analysis of the time series of historical hurricane
    numbers: two approaches we have taken in the past are described in~\citet{j88} or \citet{j90}.
    \item Time series analysis of sea surface temperature (SST), combined with analysis of the
          relationship between SST and hurricane numbers (see~\citet{j91})
    \item Analysis of the relationships between other variables and
          hurricane numbers (for instance, see~\citet{blakeg04})
    \item Use of numerical prediction models of the ocean and/or atmosphere
\end{itemize}

We are pursuing a number of lines of analysis related to these different possible
methods. In this study, we consider again what can be achieved with the first, and simplest approach, of time
series analysis of the hurricane number time series. In particular, we consider the method described
in~\citet{j90}, but now for intense hurricanes.
In that paper, we made the assumptions that the number of hurricanes is a
poisson variable, and that the poisson rate was constant from 1900 to
1994 and then jumped up to a new level for the period 1995-2006.
Based on these assumptions, we worked out how best to predict 2006, where
\emph{best} is defined as minimising RMSE.

The answer was that the best predictions for 2006 could be made by combining the
average number of hurricanes per year during the long baseline (1900-2005)
with the average numbers of hurricanes per year during the short recent period (1995-2005)
in the proportions 39\% to 61\%.
This result may seem slightly counterintuitive, since the earlier data is not really representative
of the climate of 2006, but it can be explained as follows.
A prediction based on the recent data suffers the problem that there are only 11 values on which
to calculate the average. Although such a prediction is unbiased, it has a high variance.
A prediction based on the average of the whole data (1900-2005), on the other hand,
will be biased, because the rates for some of the earlier data are low relative to the current climate,
but will have a lower variance.
Finally a prediction based on the optimal combination of the two periods can beat either of these
simpler predictions, and the proportions given above are estimates of the optimal proportions
based on a plug-in estimator.

We now extend this analysis to look at \emph{intense} hurricanes.
The question is: should the mixing of long and short baselines be the same for estimating numbers of
intense hurricanes, or should it be different, and, if it is different, what should it be?
Intuition might suggest that, since intense hurricanes are rarer, we should take relatively more account
of the earlier data. Is this intuition correct?
In section 2 we consider, in a simple idealised setting, how we would expect the optimal mixing to change
as we move to more intense hurricanes, assuming that the ratios of the numbers of hurricanes of
different intensities in the long and short baselines don't change.
In section 3 we then derive estimates of the optimal mixing proportions for real hurricane data for
intense hurricane numbers. Finally in section 4 we summarise and discuss our results.

\section{Optimal predictions for intense hurricanes}

We set up the hurricane prediction problem as follows.
We define two poisson distribution populations as:

\begin{eqnarray}
Y_{1,j} \sim \mbox{Pois}(\lambda_i), \ \ j=1,\ldots,n_1 \\\
Y_{2,j} \sim \mbox{Pois}(\lambda_i), \ \ j=1,\ldots,n_2
\end{eqnarray}

We take $Y_1$ to be the numbers of hurricanes per year for the recent 11 year period from
1995 to 2005,
and $Y_2$ to be the numbers of hurricanes per year for the earlier 95 year period, from
1900 to 1994.
Our interest lies in predicting the value ($Y_{1,n_1+1}$)
and the expected value ($E(Y_{1,n_1+1})$)
of a new observation from
the first population, where $n_1+1$ is the year 2006.

To make predictions we use various sample means, defined as:

\begin{equation}
\hat{\lambda}_i = \frac{1}{n_i} \sum_{j=1}^{n_i} Y_{i,j}.
\end{equation}

The three predictors we consider are the sample mean of the recent 11 year period,
the sample mean of the whole 106 year period (from 1900 to 2005), and an optimal combination of the two.
We write these three predictors as:

\begin{eqnarray}
\hat{Y}_{1,n_1+1} &=& \hat{\lambda}_1 \\
Y_{1,n_1+1}^\dagger &=& \frac{n_1}{n_1+n_2} \hat{\lambda}_1 + \frac{n_2}{n_1+n_2} \hat{\lambda}_2\\
Y_{1,n_1+1}^*(\alpha) &=& \alpha \hat{\lambda}_1 + (1-\alpha) \hat{\lambda}_2
\end{eqnarray}

In~\citet{j90} we showed that the optimal mixing parameter $\alpha$ in the third of these
predictors can be estimated by:
\begin{equation}\label{alpha}
\hat{\alpha}
= \frac{n_1 n_2 (\lambda_2-\lambda_1)^2 + n_1 \lambda_2}{n_1 n_2 (\lambda_2-\lambda_1)^2
                                        + n_2 \lambda_1 + n_1 \lambda_2}
\end{equation}

We now consider how $\alpha$ would change as the rates change.
If we replace both $\lambda$'s with adjusted rates $r \lambda$
(thus keeping the ratio of numbers between the two baselines constant,
but changing the absolute numbers in each baseline period)
then we get:

\begin{eqnarray}\label{alphar}
\hat{\alpha}
&=& \frac{n_1 n_2 (r \lambda_2-r \lambda_1)^2 + n_1 r \lambda_2}{n_1 n_2 (r \lambda_2-r \lambda_1)^2
                                        + n_2 r\lambda_1 + n_1 r\lambda_2}\\
&=& \frac{n_1 n_2 r (\lambda_2-\lambda_1)^2 + n_1 \lambda_2}{n_1 n_2 r (\lambda_2-\lambda_1)^2
                                        + n_2 \lambda_1 + n_1 \lambda_2}
\end{eqnarray}

The $r$'s do not cancel, suggesting that the optimal mixing proportions $\alpha$ will depend on the
frequency of the events.

In the limit as $r$ tends to zero this gives:
\begin{eqnarray}\label{alpha0}
\hat{\alpha} &=& \frac{n_1 \lambda_2}{n_2 \lambda_1 + n_1 \lambda_2}
\end{eqnarray}

In other words, however rare the event we should never completely ignore the data from the recent category.

In the limit as $r$ tends to infinity this gives:
\begin{eqnarray}\label{alpha1}
\hat{\alpha} &=& 1
\end{eqnarray}

In other words, for very frequent events we can ignore the earlier data, and rely solely on
the data in the recent category.

\begin{figure}[!hb]
  \begin{center}
    \scalebox{0.7}{\includegraphics{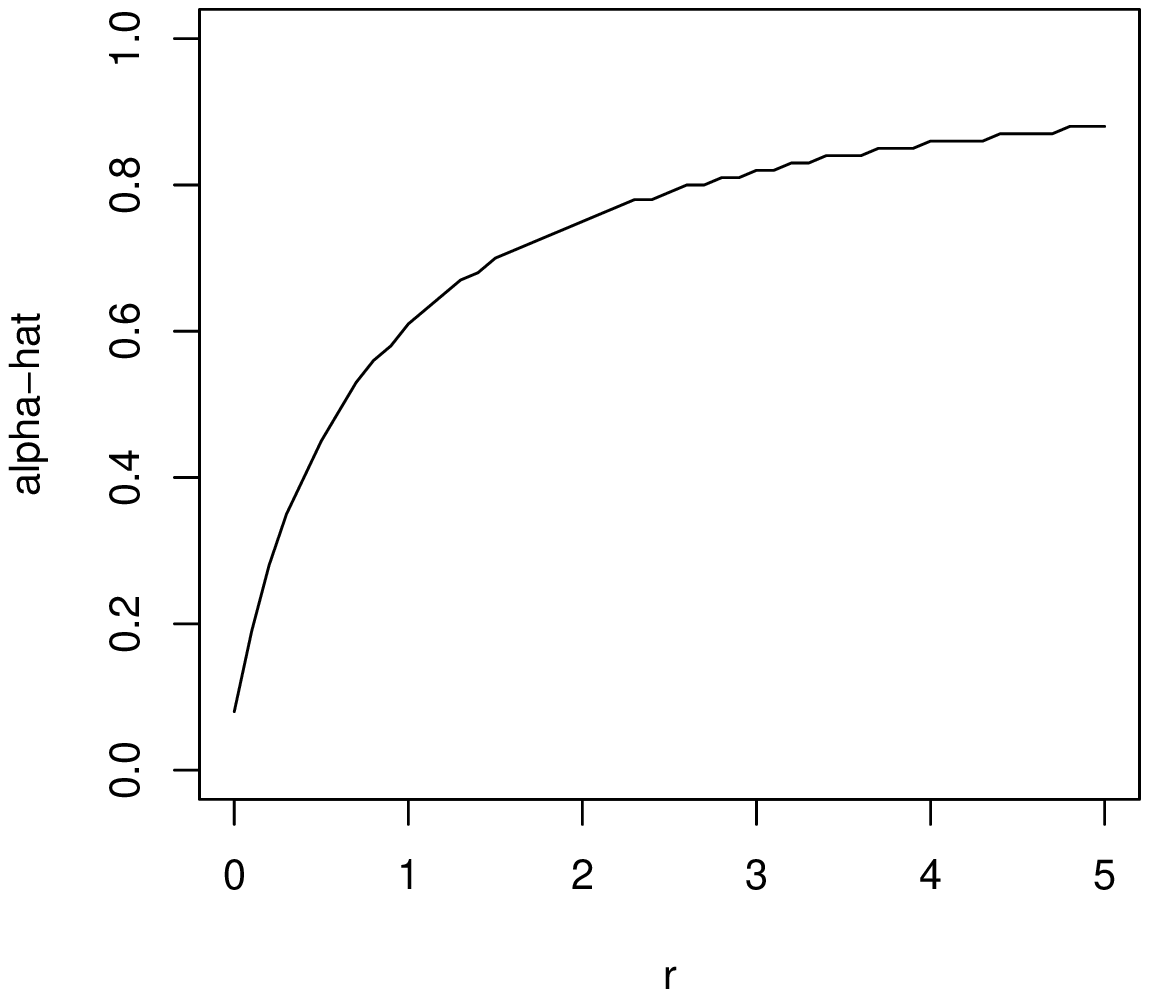}}
  \end{center}
    \caption{
The variation in the optimal mixing parameter $\alpha$ as the poisson
intensity of the modelled processes varies (based on equation~\ref{alphar}).
     }
     \label{f01}
  \begin{center}
    \scalebox{0.7}{\includegraphics{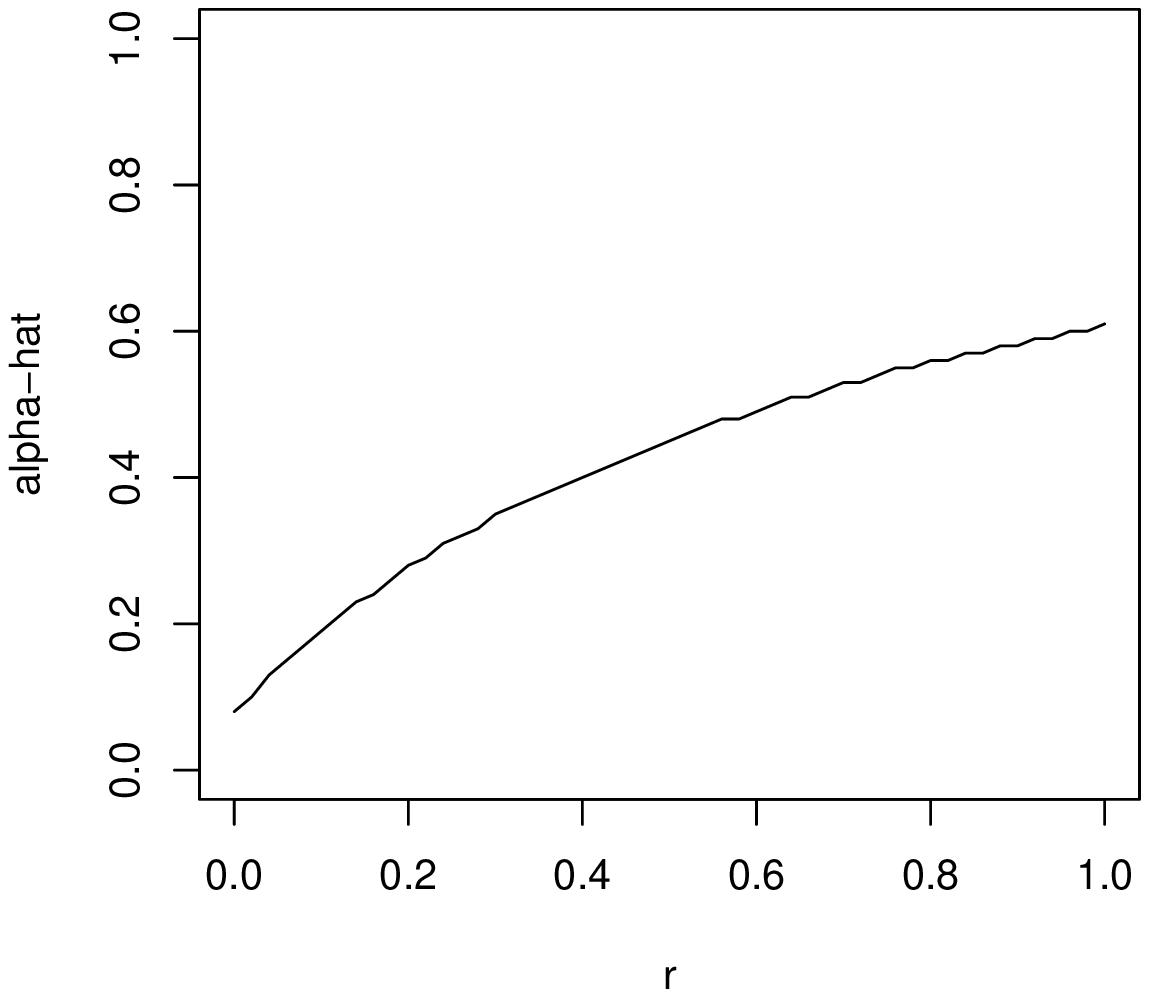}}
  \end{center}
    \caption{
Same as above for a close-up of the horizontal axis.     }
     \label{f02}
\end{figure}

For the example given in~\citet{j90} (from which we take the values of $n1, n2, \lambda_1$ and $\lambda_2$)
the first of these limits gives a value
of 0.08 i.e. 8\%, and the variation of $\alpha$ for values of $r$ between $0$
and $2$ is given in figure~\ref{f01}.
We see that for low values of $r$, $\alpha$ is very low. Thus, for very rare
events we should put most emphasis on the long baseline. As $r$ increases and
events become more common we can put more emphasis on the recent data, and less on the earlier data.

\section{Example}

We now estimate alpha for five cases, in which we consider
progressively more and more intense hurricanes. The categories are as follows:
\begin{itemize}
    \item Case 1: hurricanes of category 1-5
    \item Case 2: hurricanes of category 2-5
    \item Case 3: hurricanes of category 3-5
    \item Case 4: hurricanes of category 4-5
    \item Case 5: hurricanes of category 5
\end{itemize}

The numbers of hurricanes in each category for the two periods
1900-1994 and 1995-2005 are given in columns 2 and 3 of table 1.
Columns 4 and 5 give the same values as average annual rates, and column 6
gives the ratio of these average annual rates (95-05 divided by 00-94).
Column 7 then gives the estimated value of $\alpha$, based on these numbers.

\begin{table}[h!]
  \centering
\begin{tabular}{|c|c|c|c|c|c|c|}
 \hline
 1 & 2 & 3 & 4 & 5 & 6 & 7\\
 \hline
 cat&  00-94&   95-05& 00-94 & 95-05 & ratio & $\hat{\alpha}$\\
 \hline
1-5& 157&  25&   1.653&   2.273&   1.375&   0.660\\
2-5&  94&  17&   0.989&   1.545&   1.562&   0.695\\
3-5&  59&  10&   0.621&   0.909&   1.464&   0.520\\
4-5&  16&   1&   0.168&   0.091&   0.540&   0.485\\
  5&   3&   0&   0.032&   0.000&   0.000&   1.000\\
 \hline
\end{tabular}
\caption{The columns show:
(1) the row number,
(2) the number of landfalling hurricanes in these category between 1900 and 1994,
(3) the number of landfalling hurricanes in these categories between 1995 and 2005,
(4) the average annual number of landfalling hurricanes between 1900 and 1994,
(5) the average annual number of landfalling hurricanes between 1995 and 2005,
(6) the ratio of column 5 to column 4 (giving the change in the average annual number from the earlier
period to the recent period, and
(7) the value of the mixing coefficient $\alpha$.
}
\end{table}

For categories 1-5, 2-5 and 3-5 we see that the average annual rates increase
when moving from the early period to the recent period (compare columns 4 and 5). Only for categories 4-5 and 5
do they decrease. There have been no category 5 events at all in the recent period.
The values of $\alpha$ for categories 1-5 and 2-5 are very close, although 2-5 is slightly
higher, which is perhaps surprising given the discussion in section 2 above, which suggested
that $\alpha$ should decrease as events become rarer.
\footnote{Note that the value of $\alpha$ for cat 1-5 doesn't quite agree with the value given
in the previous paper, even though one might expect it should. The difference arises because the
results given here are based on the 2005 Hurdat database, rather than the 2004 Hurdat database.
Comparing these two databases, the number of hurricanes in the period 1900-1994 has increased
by one from 156 to 157. We also incorrectly guessed that 2005 would have 4 landfalling hurricanes,
when in fact it had 5.}
The reason for this increase in $\alpha$ is that the increase in the number of cat 2-5 hurricanes
is a larger percentage increase than the increase in the number of cat 1-5 hurricanes.
In other words there have been disproportionately more hurricanes of cat 2 and above in the recent period.
In the cat 3-5 and cat 4-5 cases, the value of $\alpha$ is lower, as we'd expect, just because
there are fewer of these events. The cat 5 case is interesting. The estimate of $\alpha$ is 1, suggesting
that we should put all our weight on the recent period, in which there have been no hurricanes.
This is clearly not sensible, and in fact the model is breaking down at this point: the problem is that the model can't estimate
hurricane rates from data in which no hurricanes occurred. Estimating rates for cat 5 hurricanes
should clearly be approached differently.

Using the values of $\alpha$ in table 1 we can make predictions for 2006 of the numbers
of hurricanes in each of these categories, as well as making the simpler
predictions based purely on averages of hurricane numbers over the long and short
baselines. These are shown in table 2.
Column 2 shows the long baseline prediction, column 3 shows the short baseline
prediction (which is the same as column 5 in table 1), and column 4 shows the optimal prediction based on the value
of $\alpha$. Columns 5, 6 and 7 show the same results, but now as a fraction
of the total number of hurricanes.
Columns 8 and 9 show the estimated sizes of errors on the optimal predictions,
first as absolute numbers (column 8), and then as a percentage of the optimal prediction.

\begin{table}[h!]
  \centering
\begin{tabular}{|c|c|c|c|c|c|c|c|c|}
 \hline
 1 & 2 & 3 & 4 & 5 & 6 & 7 & 8 & 9\\
 \hline
 cat& long bl& short             bl& optimal& long bl& short bl            & optimal& rmse2 & rmse2 (\%)\\
 \hline
1-5&   1.717&   2.273&   2.062& 100.000& 100.000& 100.000&   0.369&  17.9\\
2-5&   1.047&   1.545&   1.376&  60.989&  68.000&  66.708&   0.312&  22.7\\
3-5&   0.651&   0.909&   0.771&  37.912&  40.000&  37.378&   0.207&  26.9\\
4-5&   0.160&   0.091&   0.131&   9.341&   4.000&   6.344&   0.063&  48.4\\
  5&   0.028&   0.000&   0.000&   1.648&   0.000&   0.000&   0.000& NaN  \\
 \hline
\end{tabular}
\caption{The columns show
(1) the row number,
(2) a prediction for the number of hurricanes in 2006 based on the average number of hurricanes
per year during the period 1900-2005,
(3) a prediction for the number of hurricanes in 2006 based on the average number of hurricanes
per year during the period 1995-2005,
(4) the optimal combination of the predictions in columns 2 and 3 using the $\alpha$ from the last column
in table 1,
(5), (6) and (7): the predictions in columns 3,4 and 5 expressed as a percentage of the total number of hurricanes,
(8) the absolute error estimated for the prediction in column 4, and
(9) the percentage error estimated for the prediction in column 9.
}
\end{table}

What can we see in these results? As we'd expect, the optimal predictions
(in absolute number terms) lie between the long and short baseline predictions.
The cat 3-5 and cat 4-5 predictions lie closer to the long baseline predictions
than the cat 1-5 and cat 2-5 predictions because of the lower value of $\alpha$. The cat 5 predictions are included for completeness,
but should be ignored because of the model failure described above.

The percentage results (columns 5 to 7) give a little more colour. The prediction of the proportion of hurricanes
that are cat 2-5 has gone up relative to the long baseline. The prediction of the proportion
of hurricanes that are cat 4-5 has gone down relative to the long baseline, but not as far down
as one would derive from the short baseline. Finally, looking at the errors on these predictions,
we see that the numbers of more intense hurricanes are estimated more accurately in an absolute
sense but less accurately in a relative sense. The relative errors on the number of cat 4-5 hurricanes is between
2.5 and 3 times the error on the number of all hurricanes.

\section{Discussion}

We have considered how to apply the baseline mixing model of~\citet{j90} to the case of intense hurricanes.

We have shown at a theoretical level that for more intense hurricanes we would expect the
method to put more weight on earlier data. This is simply because intense hurricanes are less
frequent.
Using real historical data, we find that weights of 67\% to 33\% for cat 1-5 hurricanes
(for short baseline and long baseline forecasts
respectively) become weights of 52\% and 48\% for cat 3-5 hurricanes.

We have also found a flaw in the model, which occurs for the cat 5 hurricane case.
In this case, there have been no landfalling hurricanes at all in the recent period.
The model then (wrongly) puts 100\% weight on the recent data. Clearly this result should
not be trusted, since it seems very unlikely that we can conclude that cat 5 hurricanes
have become impossible just because we haven't seen one in the last 11 years.
At a fundamental level this exposes a flaw with the idea of fitting the poisson distribution with
a single parameter estimate to model
hurricane numbers. One way to resolve this problem is to use Bayesian statistics. We plan to
investigate this further in another article.

\section{Legal statement}

SJ was employed by RMS at the time that this article was written.
However, neither the research behind this article nor the writing
of this article were in the course of his employment, (where 'in
the course of their employment' is within the meaning of the
Copyright, Designs and Patents Act 1988, Section 11), nor were
they in the course of his normal duties, or in the course of
duties falling outside his normal duties but specifically assigned
to him (where 'in the course of his normal duties' and 'in the
course of duties falling outside his normal duties' are within the
meanings of the Patents Act 1977, Section 39). Furthermore the
article does not contain any proprietary information or trade
secrets of RMS. As a result, the authors are the owners of all the
intellectual property rights (including, but not limited to,
copyright, moral rights, design rights and rights to inventions)
associated with and arising from this article. The authors reserve
all these rights. No-one may reproduce, store or transmit, in any
form or by any means, any part of this article without the
authors' prior written permission. The moral rights of the authors
have been asserted.

The contents of this article reflect the authors' personal
opinions at the point in time at which this article was submitted
for publication. However, by the very nature of ongoing research,
they do not necessarily reflect the authors' current opinions. In
addition, they do not necessarily reflect the opinions of the
authors' employers.

\bibliography{chp_intense}

\begin{thebibliography}{4}
\providecommand{\natexlab}[1]{#1}
\providecommand{\url}[1]{\texttt{#1}}
\expandafter\ifx\csname urlstyle\endcsname\relax
  \providecommand{\doi}[1]{doi: #1}\else
  \providecommand{\doi}{doi: \begingroup \urlstyle{rm}\Url}\fi

\bibitem[Blake and Gray(2004)]{blakeg04}
E~Blake and W~Gray.
\newblock {Prediction of August Atlantic basin hurricane activity}.
\newblock \emph{Weather and Forecasting}, 19:\penalty0 1044--1060, 2004.

\bibitem[Jewson et~al.(2005)Jewson, Casey, and Penzer]{j90}
S~Jewson, C~Casey, and J~Penzer.
\newblock {Year ahead prediction of US landfalling hurricane numbers: the
  optimal combination of long and short baselines}.
\newblock \emph{arxiv:physics/0512113}, 2005.

\bibitem[Khare and Jewson(2005)]{j88}
S~Khare and S~Jewson.
\newblock {Year ahead prediction of US landfalling hurricane numbers: intense
  hurricanes}.
\newblock \emph{arxiv:physics/0512092}, 2005.

\bibitem[Meagher and Jewson(2006)]{j91}
J~Meagher and S~Jewson.
\newblock {Year ahead prediction of hurricane season SST in the tropical
  Atlantic}.
\newblock \emph{arxiv:physics/0606185}, 2006.

\end{thebibliography}

%%%%%%%%%%%%%%%%%%%%%%%%%%%%%%%%%%%%%%%%%%%%%%%%%%%%%%%%%%%%%%%%%%%%%%%%%%%%%%%%%%%%%%%%%

\newpage

\end{document}